\documentstyle[prd,aps,epsfig]{revtex}
\tighten
\begin{document}
\draft
\twocolumn[\hsize\textwidth\columnwidth\hsize\csname
@twocolumnfalse\endcsname
\title{BOUND STATES IN MONOPOLES: SOURCES FOR UHECR ?}
\author{Eric HUGUET$^1$ and Patrick PETER$^2$}
\address{$^1$D\'epartement d'Astrophysique Stellaire et Galactique,\\
Observatoire de Paris-Meudon, 92195 Meudon, France,\\
and Universit\'e Paris VII, place Jussieu, 75005 Paris, France, \\
$^2$D\'epartement d'Astrophysique Relativiste et de Cosmologie,\\
Observatoire de Paris-Meudon, UPR 176, CNRS, 92195 Meudon, France. \\
{\rm Email: eric.huguet@obspm.fr,patrick.peter@obspm.fr}}
\date{\today}
\maketitle
\begin{abstract}
Bound states in monopoles are studied through a simplified,
Witten-like model. As the overall structure is determined in full
details, it is shown in particular that only those states having a
vanishing angular momentum are allowed; for those, the energy spectrum
is derived numerically and an approximation is set up that allows an
easy description in terms of wavefunctions, usefull for further
applications. The monopoles are then proposed as candidates of ultra
high energy cosmic rays, an hypothesis that should soon be testable
through the Pierre Auger Observatory.
\end{abstract}

\vskip2pc
]

\section{Introduction}

Among the various topological defects that may have formed during
phase transitions in the early universe~\cite{kibble}, monopoles are
absolutely unavoidable in Grand Unified Theories (GUT), according to
which there was a time when all symmetries were unified in a single
semi-simple gauge group, since electromagnetism is an exact (unbroken)
symmetry still now. They are however usually considered as a
sufficient nuisance~\cite{preskill} (in the sense that they would very
rapidely come to dominate the universe) that inflation is generally
invoked right after their time of formation so that they are
completely diluted~\cite{inflation} and thus unobservable. These
results are based on the by-now very standard view that they
originated at the GUT phase transition, i.e., at an energy around
$10^{15}-10^{16}$ GeV. Many ways out however have been proposed other
than inflationnary scenarios, among which the Langacker-Pi
mechanism~\cite{pi} which relies upon using cosmic strings to connect
the monopoles and anti-monopoles pairs, thereby effectively enhancing
considerably the decay probability, thus reducing the remnant monopole
density. Although the model, in its original presentation, suffers
from many drawbacks, it has at least the advantage of proposing
another solution, not involving inflation and letting open other
alternative possibilities. Domain walls have also been used to sweep
them away~\cite{sweep}; in all cases, undesirable monopoles are gotten
rid of by means of higher dimensional topological defects~\cite{book}.

Yet another alternative possible solution to the monopole excess
problem is the simplest one, although completely overlooked until
recently: it consists in noting that the monopole density is in fact
proportional to the fourth power of the energy scale $\eta$ at which
the symmetry breaking during which monopoles were generated took
place, given the monopole mass $m_{_M}$ is essentially $\eta$ times
the inverse of the corresponding coupling constant, i.e., $\sim 137$
in the case of electromagnetism (as ought to be the case):
\begin{equation} \Omega_{_M} h^2 \simeq 10^{11} \left(
{\eta\over 10^{14}\hbox{GeV}} \right)^3 \left( {m_{_M} 
\over 10^{16} \hbox{GeV}}\right) ,\end{equation}
(here we note $\Omega_{_M}$ the monopole density in units of the
critical closure density, and $h$ the Hubble constant in units of 100
km~$\cdot$~s$^{-1}\cdot$~Mpc$^{-1}$) so that it suffices to lower
$\eta$ to $\alt 10^9$~Gev in order to cure the density problem. This
is the solution we shall adopt here, for it might also provide a
useful explanation for the high energy cosmic ray
mystery~\cite{kp}. (Note in this regard that one can also simply
assume the monopole overdensity problem to be cured somehow and
investigate anyway the possibility that they give rise to high energy
cosmic rays~\cite{bs}.) This way, the possibility that monopoles are
still present in the universe is still reasonnable.

Once the mass scale is fixed, the relevant physics still needs be
properly clarified in order to enable one to study the interactions
between monopoles and other particles. In this regard, models have
been suggested where bound states of scalar~\cite{scalar}, gauge
vector bosons~\cite{vector} or fermions~\cite{fermion} can form in
SO(3) t'Hooft-Polyakov monopoles~\cite{tP}. Those modify the
scattering solutions and can enhance greatly the cross sections by
means of the Callan-Rubakov effect~\cite{cr}. It is the same kind of
approach we wish to present here, although using a simplified Witten
like model (generally used to describe current-carrying cosmic
strings~\cite{witten,neutral} or domain walls~\cite{wall}) and a
completely different method for calculating the bound state energy
levels. Our model, contrary to the other proposed models having bound
states, presents the advantage (apart from its simplicity) that the
bound states are present for dynamical reasons, i.e., they do not
exist for all values of the underlying parameters. This means that we
can consider that particles may be trapped in the monopole even though
they might belong to representations that are not directly related to
that of the monopole fields. Besides, we shall work using the
Julia--Zee fully numerical solution~\cite{jz} and not the explicit
Bogomol'nyi--Prasad--Sommerfield (BPS) analytic limit~\cite{ps}.

In what follows, we shall first define our model and show that bound
states may be present in monopoles, provided their angular momentum
vanish, exactly as in the Callan-Rubakov mechanism~\cite{cr}. This
not-quite-obvious result follows from the requirement that bound
states do exist in the monopole, which in turn seriously constrains
the possible couplings between the fields involved.

Having explored the full microscopic structure of a monopole having
bound states, we go on to investigate the possibility that, in a
fashion similar to Witten superconducting cosmic string
models~\cite{witten}, through the existence of vortons~\cite{vortons}
(rotating superconducting cosmic string loops configurations), these
monopoles could be the source of Ultra High Energy Cosmic Rays
(UHECR). Their acceleration, once they are coupled to
electromagnetism, poses no particular problem as long as an external
magnetic field exists in the accelerating region under
consideration. Similarly to the model based on vortons~\cite{bona},
they can propagate along huge (cosmologically speaking) distances, and
thus have the ability to reach us fairly easily. Their interaction
cross-section with air nuclei can be obtained through the
Callan-Rubakov effet~\cite{cr} and is thus evaluated to be hadronic in
nature. The expected properties of the resulting UHECR distribution
are essentially those derived in the vorton case~\cite{bona} and can
therefore reproduce the existing data. More data, thanks for instance
to the Pierre Auger Observatory project~\cite{PAO}, will give a
definite answer concerning this possibility.

\section{The Monopole State}

The model we shall use in what follows involves the symmetry breaking
of an SO(3) invariance by means of a Higgs field $\Phi$ belonging to
the {\bf 3} representation of SO(3), coupled to a complex scalar field
$\Sigma$ which we assume, for the sake of simplicity, not to be
coupled to the gauge field ${\bf A}^\mu$ of SO(3). This assumption
should of course be modified when one wants to evaluate the long-range
electromagnetic interaction of the resulting monopole with other
particle, but we shall show latter on how this can be achieved. With a
metric convention having positive signature, we have the model
\begin{eqnarray} {\cal L} &=& -{1\over 2} (D_\mu \Phi ^a)^\dagger (D^\mu
\Phi_a) -{1\over 2} (\partial_\mu \Sigma)^\star (\partial^\mu \Sigma )
-{1\over 4} F^a_{\mu\nu} F_a^{\mu\nu} 
\nonumber \\
& & -{\lambda_\phi \over 4} (\Phi_a \Phi^a -\eta^2)^2 - f (\Phi_a
\Phi^a -\eta^2) |\Sigma |^2 \nonumber \\
& & - {m_\sigma^2\over 2} |\Sigma |^2 -
{\lambda_\sigma \over 4} |\Sigma |^4,
\label{lag}\end{eqnarray}
where the covariant derivative is
\begin{equation} D_\mu \Phi _a \equiv \partial_\mu \Phi_a
-q\varepsilon _{ab} ^{\ \ c} A_\mu ^b \Phi _c,\end{equation}
and the gauge field strength
\begin{equation} F_{\mu \nu} ^a \equiv \partial _\mu A^a_\mu -
\partial _\nu A^a_\mu + q\varepsilon ^a_{\ bc} A^b_\mu
A^c_\nu.\end{equation} A static monopole configuration then has the
form~\cite{tP,jz}, in spherical coordinates $x^i \equiv (r, \theta,
\phi)$
\begin{equation} \Phi^a = \eta h(r) {x^a \over r}
,\label{phia}\end{equation} 
\begin{equation} A^a_i = - {1-K(r) \over q r^2} \, \varepsilon ^a_{\ ij}
x^j, \ \ \ A^a_0 = 0.\label{Amua}\end{equation} The field equations
for the configuration~(\ref{phia}) and~(\ref{Amua}), with the $\Sigma$
field not taken into account, i.e. for the ordinary 't~Hooft--Poyakov
monopole, read
\begin{equation} {1\over r^2} {d\over dr} (r^2 {dh\over dr}) = {2\over
r^2} h K^2 + \lambda_\phi \eta^2 h (h^2 -1)\end{equation}
\begin{equation} {d^2K\over dr^2} = q^2 \eta^2 Kh^2 +{1\over r^2}
K(K^2-1),\end{equation}
with boundary conditions
\begin{equation} h(0)=K(\infty)=0,\ \ \ h(\infty) = K(0)
=1.\end{equation} 

\begin{figure}
\centering
\epsfig{figure=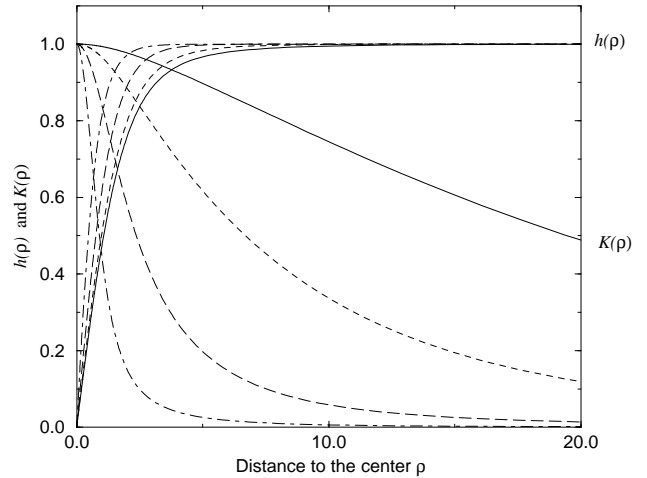, width=9cm}
\caption{
Amplitude of the field functions $h(\rho)$ and $K(\rho)$ vs the
rescalled unit of length $\rho = r\lambda_\phi^{1/2}\eta$. Curves are
shown for $q^2/\lambda_\phi = 10^{-2}$ (solid line), $0.1$ (dashed
line), $1$ (long dashed line) and $10$ (dot-dashed
line).}\label{bckgrdfields}
\end{figure}

These equations have been solved numerically for
various values of the only relevant parameter $q^2/\lambda_\phi$ and
produce the characteristic curves on Figure~\ref{bckgrdfields}, where
$h$ and $K$ are shown as functions of the rescalled unit of length
$\rho = r\sqrt{\lambda_\phi} \eta$.

\begin{figure}
\centering
\epsfig{figure=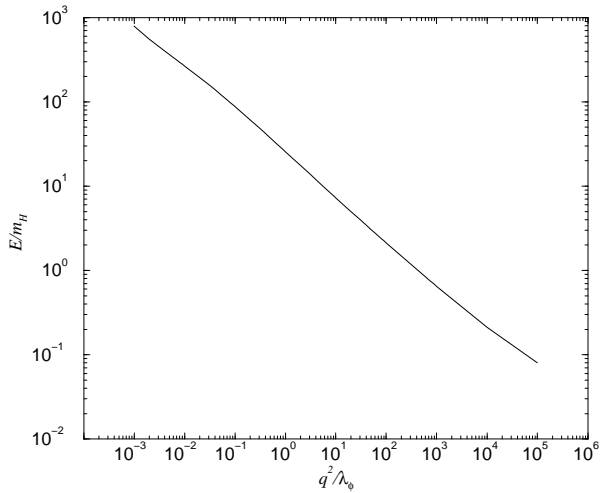, width=9cm}
\caption{Variation of the total monopole energy as a function of
$q^2/\lambda_\phi$.}
\label{energyq2}
\end{figure}

The total energy of the monopole is expressible simply in terms of
$q^2/\lambda_\phi$ and the Higgs field mass $m_{_H} =
\sqrt{\lambda_\phi} \eta$ as
\begin{eqnarray} E_{_M} &=& 4\pi m_{_H} \int \rho d\,\rho \Big\{ {K'^2
\over (q^2/\lambda_\phi) \rho^4} + {1\over 2} h'^2 + {K^2 h^2\over
\rho^2} \nonumber \\ &+& {1\over 4} (h^2-1)^2 + {(1-K^2)^2 \over
2(q^2/\lambda_\phi)\rho^4}  \Big\} ,\end{eqnarray} which
is shown on Figure~\ref{energyq2} as a function of $q^2/\lambda_\phi$.

\section{The Scalar Condensate}

\begin{figure}
\centering
\epsfig{figure=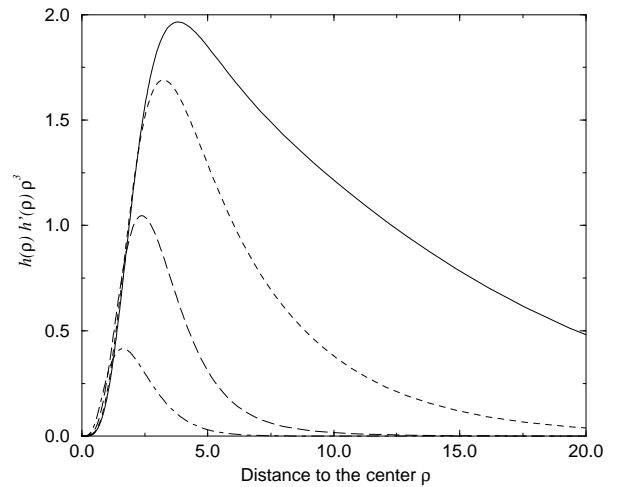, width=9cm}
\caption{Value of the function $\rho^3 h(\rho) dh/d\rho$ which
determines the possibility of bound states for nonzero values of the
angular momentum $\ell$ vs the rescalled unit of length $\rho =
r\lambda_\phi^{1/2}\eta$. As on Figure~\ref{bckgrdfields}, curves are
shown for $q^2/\lambda_\phi = 10^{-2}$ (solid line), $0.1$ (dashed
line), $1$ (long dashed line) and $10$ (dot-dashed line).}
\label{bckgrdcond}
\end{figure}

In this background, we now investigate the behaviour of the bosonic
field $\Sigma$ by first looking at the field equation in which the
nonlinar term is omitted. Separating space and time variables in the
form
\begin{equation} \Sigma (x^\mu) = \sigma (r) \hbox{e}^{i\omega t}
Y_{\ell m} (\theta,\phi),\label{sigma}\end{equation} the field
equation for $\Sigma$ in the monopole background gives the
Schr\"odinger-like eigenvalue equation for the amplitude of this field
as
\begin{equation} -\Delta \sigma + V(r) \sigma = \omega^2
\sigma,\label{eigen}\end{equation}
where
\begin{equation} V(r)= [{\ell (\ell +1)\over r^2}+ 2f\eta^2
(h^2-1) + m_\sigma^2 ] \end{equation}
which will admit bound state eigensolutions ($\omega^2 < m_\sigma^2$) provided
that the potential $V$ satisfies either
\begin{equation} V < 0 \Longrightarrow \ell =0 \ \ \& \ \ m_\sigma^2 < 2 f\eta
^2,\end{equation} or
\begin{equation} \exists R \in [0,\infty [ \ \ ; \ \ {dV\over
dr}\Big|_{r=R} = 0 \ \ \hbox{and} \ \ V(R) <0.\end{equation}
In the latter case, denoting by a prime a derivative with respect to
the rescalled distance $\rho$, one finds that the minimum of the
potential would be for $\rho$ such that
\begin{equation} {\lambda_\phi\over 2 f} \ell (\ell +1) = h h' \rho^3.
\label{cond}\end{equation}
On Figure~\ref{bckgrdcond} is plotted the right hand side of this
relation against $\rho$ for three orders of magnitude of the relevant
underlying parameter $q^2/\lambda_\phi$, and it is seen that in
general this function is of order unity throughout its range of
variation. Hence, in order for the condition~(\ref{cond}) to be
satisfied, it is necessary that the left hand side be also of order
unity. Assuming $\lambda_\phi$ and $f$ to have comparable values (they
are both quartic interaction term coupling constants), this leads to
the constraint
\begin{equation} \ell (\ell +1) \sim 1,\end{equation}
which shows that no bound state is expected for large angular
momentum. For this reason, we restricted our analysis to vanishing
angular momentum states $\ell =0$.

Yet another way to convince oneself that the bound states should be
restricted to those with vanishing angular momentum consists in
investigating the full, non perturbative, field equation for $\Sigma$,
still within the separated form~(\ref{sigma}), that is, assuming the
resulting state to be still an eigenstate of the angular momentum. The
existence of the non linear term would then imply $|Y_{\ell m}|^2$ to
be a constant, i.e., independent of both angular variables $\theta$
and $\phi$. Thus, only the case $\ell =0$ can satisfy the non linear
equation, so that the background $\sigma$ field has vanishing angular
momentum.

\begin{figure}
\centering
\epsfig{figure=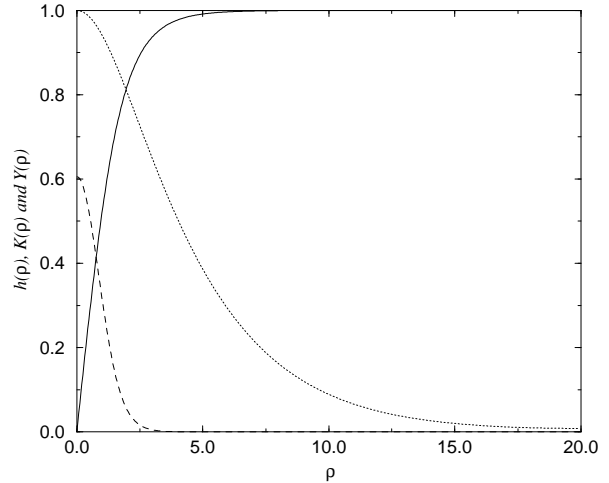, width=9cm}
\caption{Fields function $h(\rho)$ [full line], $K(\rho)$ [dotted
line] and $Y(\rho)$ [dashed line] for the set of parameters
$\alpha_1=0.1$, $\alpha_2=0.6$ and $\alpha_3=0.6$.}
\label{vacuum}
\end{figure}

We now therefore consider again the ansatz~(\ref{sigma}) with no
spherical harmonics included and in the region of parameter space
where a condensate might form, i.e. we assume $m_\sigma ^2 < 2 f
\eta^2$. Setting $\omega =0$ yields the actual vacuum state as the
solution of the field equations
\begin{eqnarray} & &{1\over r^2} {d\over dr} (r^2 {dh\over dr}) = {2\over
r^2} h K^2 + \lambda_\phi \eta^2 h (h^2 -1) + 2 f \eta^2 h \sigma^2,\\
& & {d^2K\over dr^2} = q^2 \eta^2 K h^2 +{1\over r^2} K(K^2 - 1),\\ 
& & {1\over r^2} {d\over dr} (r^2 {d\sigma \over dr}) = [
m_\sigma^2 + 2 f\eta^2 (h^2 -1)]\sigma + \lambda_\sigma
\sigma^3,\end{eqnarray}
with boundary conditions for $\sigma$ as
\begin{equation} {d\sigma \over dr} (0) = 0 , \ \ \ \lim_{r\to\infty}
\sigma =0.\end{equation}
Rescalling the field $\sigma$ through
\begin{equation} Y(\rho)= \sqrt{\lambda_\sigma} {\sigma \over
m_\sigma} \end{equation}
and defining the dimensionless parameters as
\begin{equation} \alpha_1 = {m_\sigma^2 \over \lambda_\sigma \eta^2},
\ \ \alpha_2 = {fm_\sigma^2 \over \lambda_\sigma\lambda_\phi\eta^2}, \
\ \alpha_3 = {m_\sigma^4\over
\lambda_\sigma\lambda_\phi\eta^4},\end{equation}
allows a numerical calculation of the vacuum solution. Such a
solution, obtained by means of a Successive Over Relaxation
method~\cite{SOR}, is shown for a special set of parameters
$\{\alpha_i\}$ on Figure~\ref{vacuum} (here and in what follows, the
background parameter $q^2 /\lambda_\phi$ has been fixed to the
arbitrary value $0.1$).

These fields represent the vacuum state out of which the bound state
solutions can be calculated.

\section{Quantum Theory}

In order to calculate quantum effects related to the monopole
solution, we turn to the standard solitonic approach~\cite{soliton}.
We shall for now on consider the Higgs and gauge vector fields to
represent a fixed background in which the charged scalar field
$\Sigma$ evolves. This means we describes its dynamics through an
effective Lagrangian density
\begin{eqnarray} {\cal L}_{\Sigma} & = & -{1\over 2} |\partial_\mu
\Sigma |^2 - f (\Phi_a \Phi^a - \eta^2) |\Sigma |^2\nonumber\\
& & -{m_\sigma^2\over 2}|\Sigma |^2 - {\lambda_\sigma\over 4} |\Sigma
|^4,\end{eqnarray} and the localized solutions are derivable from the
effective potential $V[\Sigma]$
\begin{equation} L_\Sigma = \int \! d^3\! x \, {\cal L}_\Sigma = \int d^3x
{1\over 2} {\partial\Sigma\over \partial t} - V[\Sigma],\end{equation}
\begin{eqnarray} V[\Sigma] &=& \int \! d^3 \! x\,  \Big\{ {1\over 2}
(\nabla \Sigma )^2 + {\lambda_\sigma\over 4} |\Sigma |^4 \nonumber \\
& & \ \ \ \ + \left[ f(\Phi_a \Phi^a - \eta^2) +{m_\sigma^2\over 2}
\right] |\Sigma |^2 \Big\}.
\end{eqnarray}
The classical solution $\Sigma_c$ derived earlier is then obtained by
minimizing $V[\Sigma]$:
\begin{equation} {\delta V\over \delta \Sigma} = 0 \Longrightarrow
\Delta \Sigma_c = [2f(\Phi_a \Phi^a - \eta^2) +m_\sigma^2] \Sigma_c +
\lambda_\sigma \Sigma_c^3,\end{equation}
where, because of the $U(1)$ symmetry in the scalar field $\Sigma$,
the later could have been chosen real.

Expanding $V[\Sigma]$ around the classical solution and noting
respectively 
\begin{equation} \tilde \eta = {1\over \sqrt{2}} (\psi +
\psi^\star)\end{equation}
and
\begin{equation} \tilde \mu = {1\over \sqrt{2}} (\psi -
\psi^\star),\end{equation}
with $\psi$ the quantum perturbation from the classical solution
$\Sigma =
\Sigma_c + \psi$, one gets
\begin{eqnarray} V[\Sigma ] & = & V[\Sigma_c] +V_{int} [\tilde \eta,
\tilde \mu]\\
& - & {1\over 2}\int \! d^3 \! x\,
\tilde \eta \Big[ {\Delta\over 2} - f (\Phi_a \Phi^a -\eta^2)
-{m_\sigma\over 2} - {3\over 2} \Sigma_c \Big] \tilde \eta \nonumber\\
& - & {1\over 2}\int \! d^3 \! x\,
\tilde \mu \Big[ {\Delta\over 2} - f (\Phi_a \Phi^a -\eta^2)
-{m_\sigma\over 2} - {1\over 2} \Sigma_c \Big] \tilde
\mu,\nonumber\end{eqnarray}
where $V_{int}[\tilde\eta,\tilde \mu]$ comprises the interaction terms
between $\tilde\eta$ and $\tilde\mu$, originating from the self
coupling $|\Sigma |^4$. These terms are not explicitely developed here
since we are only interested in the bound state solutions around the
monopole, i.e., the stationnary solutions on the basis of which the
system can be quantized. We are therefore looking for the eigenmodes
of the second derivatives of $V$ with respect to the fields
$\tilde\eta$ and $\tilde\mu$, which are then seen to satisfy the
Schr\"odinger-like equations
\begin{equation} \Big[ -\Delta +2f(\Phi^a \Phi_a -\eta^2)+m_\sigma^2 +
3\Sigma_c^2\Big]\tilde\eta_i = \omega_i^2 \tilde\eta_i,
\label{schrodeta}\end{equation}
and
\begin{equation} \Big[ -\Delta +2f(\Phi^a \Phi_a -\eta^2)+m_\sigma^2 +
\Sigma_c^2\Big]\tilde\mu_j = \Omega_j^2 \tilde\mu_j.
\label{schrodmu}\end{equation}
The eigenmodes $\tilde\eta_i$ are of two different kinds: either
$\omega_i$ belongs to a discrete set, which is the case if $\omega_i
\leq m_\sigma$, so that $\tilde\eta_i$ represents a bound state
localized around the monopole, or $\omega_i > m_\sigma$ can be
parametrized by a continuous index $q$ that can be identified with a
momentum. In the later case, the corresponding modes are diffusion,
i.e. asymptotically free states, with momentum $q$. Similar
considerations obviously apply also to the states $\tilde\mu_j$.

On the basis of this set of eigenmodes, and following Goldstone and
Jackiw~\cite{GJ}, one can in principle build the quantum theory of the
monopoles by constructing a Fock space as follows. The original
monopole state, i.e., the classical solution with no bound state, can
be boosted to aquire an arbitrary momentum $P$, thereby generating the
set $\{ |P\rangle \}$. The same can be done for monopoles with any
occupation numbers in the bound states $\{ |P,n^{(\eta)}_i,
n^{(\mu)}_j\rangle \}$. Finally, the Fock space is completed by
inclusion of the diffusion states labelled by the momenta of the
various ingoing and outgoing particles $\{ |q^{(\eta)}_1, \cdots ,
q^{(\eta)}_N, q^{(\mu)}_1, \cdots , p^{(\mu)}_M \rangle \}$. Thus, an
arbitrary quantum state belongs to the set
\begin{eqnarray} {\cal F} &=& \Big\{ |P,n^{(\eta)}_i, n^{(\mu)}_j ,
q^{(\eta)}_1, \cdots , q^{(\eta)}_N, q^{(\mu)}_1, \cdots , p^{(\mu)}_M
\rangle ,\nonumber \\
& & n^{(\eta)}_i, n^{(\mu)}_j\in {\bf N}, q^{(\eta)}_\alpha ,
q^{(\mu)}_\beta \in {\bf R}^3 \Big\}, \end{eqnarray} and this space is
assumed disconnected from the ordinary free particle space of the
theory without a monopole. Cross sections will then be calculable by
computing matrix elements of the quantum field $|\Sigma |^4$, which is
seen to involve essentially space integrals of the
eigenfunctions. Specific such calculations and their application to
primordial cosmology will be the subject of a further work, and for
now on we shall concentrate on the actual structure of both the self
potential and the bound states.

\section{The self potentials; bound states}

\begin{figure}[t]
\centering
\epsfig{figure=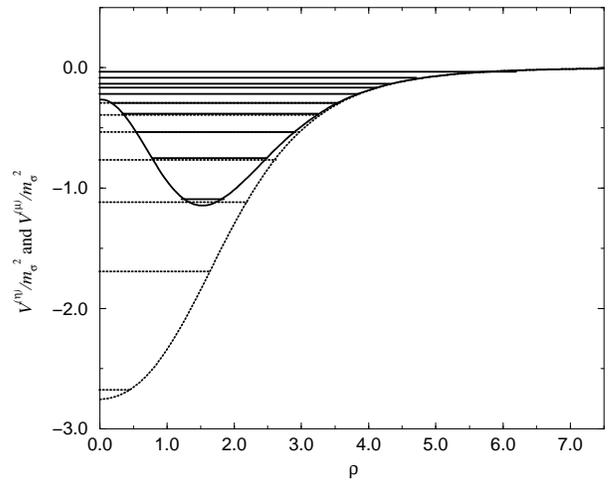, width=9cm}
\caption{Confining potential of the first [$V^{(\mu)}(r)$] and
second [$V^{(\eta)}(r)$] kind in units of the $\Sigma-$field mass
squared $m^2_\sigma$ for the set of parameters $\alpha_1 = 0.1$,
$\alpha_2 = 0.2$ and $\alpha_3 = 0.1$ as a function of the rescalled
distance to the monopole core $\rho$. In this case, the minimum of
$V^{(\mu)}(r)$ is located at $\rho=0$ while that of $V^{(\eta)}(r)$ is
for $\rho=\rho_m\not= 0$. The bound states eigenvalues
are indicated as the straight lines. The full lines correspond to
$V^{(\mu)}$ and its associated eigenvalues, while the dotted lines
stand for $V^{(\eta)}$.}
\label{pot}
\end{figure}

The purpose of this section is to exhibit the various possible
situations, depending on the underlying microscopic parameters, in
which the field $\Sigma$ can evolve. To achieve this, we rewrite the
eigenmode equations for $\tilde\eta$ and $\tilde\mu$ in the form
\begin{equation} \left\{ \matrix{[-\Delta +V^{(\eta)} (r) ]\tilde\eta
= (\omega^2 - m_\sigma^2)\tilde \eta,\cr \cr [-\Delta +V^{(\mu)} (r)
]\tilde\mu = (\omega^2 - m_\sigma^2)\tilde \mu}\right. ,\end{equation}
which define the self potentials $V^{(\eta)} (r)$ and $V^{(\mu)} (r)$
by comparison with Eqs. (\ref{schrodeta}) and (\ref{schrodmu}).  Some
characteristic shapes and amplitudes of these potentials are shown on
Figure~\ref{pot}, in rescalled units (the square of the carrier
mass), together with the corresponding eigenvalues in the same units.

The figure reveals the existence of two different kinds of self
potentials, respectively called of the first and second kind,
depending on the underlying microscopic parameters. The first kind is
characterized by an absolute minimum in the monopole core and
corresponds to a weak coupling where $f$ is small, while the second
kind, having a local maximum at $r=0$ and a minimum for some nonzero
value of the distance to the monopole core, reflects the existence of
a strong coupling. As a rule, and as could have been expected, the
energy eigenstates are more bound for confining potentials of the
first kind that for second kind. Therefore, one can expect the
lifetime of the corresponding configurations to be strongly dependent
on the parameters.

\begin{figure}
\centering
\epsfig{figure=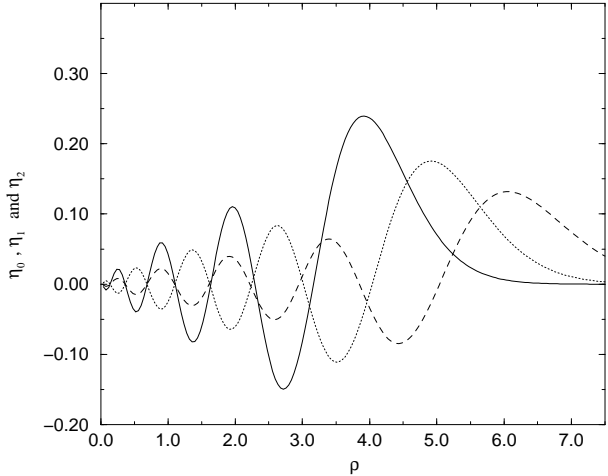, width=9cm}
\caption{An illustrative example of some wavefunctions living in a
self potential of the second kind. The set of parameters has been
chosen as on figure~\ref{pot}.}
\label{func}
\end{figure}

The figure~\ref{func} illustrates some wavefunctions living in self
potentials of the first and second kinds.

These wavefunctions are all that is needed to fully clarify the
internal structure of the monopole. The expected spectrum would have
line features as obtained by reversing the axes on Fig.~\ref{pot}.  We
therefore turn to these cross-sections before discussing the
acceleration and propagation mechanisms of monopoles.

\section{Monopole-proton interactions}

Contrary to the case of an ordinary (structureless) monopole, the
configurations we have been studying have the ability to interact
through deep-inelastic scattering with air nucleus, a proton say to
simplify matters. Indeed, a particle trapped inside the monopole core
can be scattered off the monopole, effectively ionizing it. As in the
vorton case~\cite{bona}, three possible interactions can take place,
depending on the proton energy (we keep working in the monopole
center-of-mass frame): elastic interaction, which was discussed
in~\cite{kp} and which will presumably yield an unobservable shower
accompanied with \v{C}erenkov radiation, excitation interaction
whereby a charge carrier will use the proton energy to move to a
higher energy level, and finally ionization. The last two cases are
the most important as one expects them to yield an observable signal;
they will be compared to existing data and used for making predictions
in the Pierre Auger Observatory~\cite{PAO}.

Once a trapped particle has been moved up to an excited state, it will
decay into the minimum energy state, for instance in radiating a
photon\footnote{The term ``photon'' shall be used here in the generic
sense of an unbound decay product of our $\Sigma-$field. In case
$\Sigma$ is effectively coupled to electromagnetism, this ``photon''
would be an actual photon.}, assuming there does exist such a
channel. Calling $\Delta E = E_n - E_0\sim m_\sigma$ the energy
difference between the two eigenenergies (see, e.g. Fig.~\ref{pot}),
such an observed primary photon would have energy, seen on earth
\begin{equation} E_{obs} = \gamma \Delta E,\label{energy}
\end{equation}
with $\gamma = \varepsilon / m_{_M}$ the monopole Lorentz
factor. Here, $\varepsilon$ is the total monopole energy, while
$m_{_M}\sim \eta /e$ is its mass. It should be remarked that the
monopole mass can be much larger than the carrier mass, in practice as
high as a hundred times, so that the monopole itself must be
accelerated to much higher energies than the observed $10^{20}$~eV. We
shall see however in the following section that this is not a very
serious constraint. If the energy is sufficient to ionize the
monopole, the observed energy will be also of the order of that given
on Eq.~(\ref{energy}), with $\Delta E$ simply equal to $E_0$, the
minimum energy state. In order that the reaction actually takes place,
it is necessary that the incident proton, which, as seen from the
monopole frame, is having the same Lorentz factor $\gamma$, has enough
energy to excite the corresponding state. Therefore, $\Delta E =
\gamma m_{p}$, with $m_p$ the proton mass, so that
\begin{equation} \gamma =({\varepsilon\over m_p})^{1/2} \sim
10^{6.5}.\label{g6}\end{equation}
Such a number would seem to imply states in a monopole at energies of
the order of a few thousands TeV, and therefore far below the
cosmological limit $m_{_M}\alt 10^9$ GeV. This leaves a large range of
possibilities for the ratio $\eta / m_\sigma$.

Various cases have to be taken into account if one wants to actually
understand the data in terms of bound states in monopoles. First of
all, the determination of the energy levels, as discussed above,
although specifically relevant to our case of a bosonic condensate, is
essentially the same in the case of fermionic bound states. However,
in the latter case, the Pauli exclusion principle would apply and the
overall distribution of populated states would be different: as all
bosons would preferentially occupy the same state (the lowest energy
state), fermions on the other hand would fill all the states. The
expected spectrum would therefore be qualitatively different. The
solid prediction here is the same as that for the vorton solution to
the UHECR problem~\cite{bona}, namely that of the existence of a line
spectrum, the exact form of which needing a specific model to be
determined. If such a detection was indeed achieved, the method
presented above would prove usefull to actually determine the relevant
particle physics model through a fit of the data.

Identifying the primary as a monopole is also possible through its
interactions with air nuclei. In other words, one needs to know the
monopole atom cross section, or, as emphasized above, the monopole
proton cross section. Such a calculation is feasible in principle by
making use of the quantum theory developed in Section~IV and V, and
will have to be done in case a ray spectrum is indeed observed in the
Pierre Auger Observatory~\cite{PAO}. For the time being, a rough
evaluation will be enough.

A monopole such as described here interacts with a fermionic field
such as a proton through possible exitations of the $A_0$ modes, the
so-called dyonic modes. This is a resonant exitation which effectively
confines the fermion in the core of the monopole for a long time,
making the cross-section essentially hadronic (the proton is then seen
by the monopole as a quasi-bound state). Once trapped, the proton (or
a quark therein) has time to interact with the bound states. As a
result~\cite{cr}, the interaction cross-section we are seeking is
given by:
\begin{equation} \sigma_{_{Mp}} \sim q^{-2} m_{_p}^{-2},\end{equation}
numerically of the order of a few hundreds mb. Such a large
cross-section implies that most of the monopoles arriving in the
atmosphere would be detectable, provided they carry bound states.

\section{Acceleration and propagation}

Once the internal structure of a monopole is known, it must be shown
that they have the ability to be detected as UHECR. It is the purpose
of this section to show that indeed many astrophysical sites have the
ability to accelerate them. Moreover, there is no GZK cutoff for them
since they share the vorton features of being effectively highly
charged, a fact that is compensated by their huge mass: the
electromagnetic energy losses are similar to that of a heavy nucleus
having an electric charge $Z=1/2q=137/2$~\cite{kp}, but are reduced by
inverse powers of the mass~\cite{bona}.

Accelerating a magnetic monopole in the presence of a magnetic field
is a very easy task: even the galactic magnetic field could do
it~\cite{kp} as the kinetic energy $E_{_M}$ of a monopole in a
magnetic field $B\sim 10^{-6}$~G with a coherence length $D\sim
300$~pc would be of order
\begin{equation} E_{_M} = q_{_M} B L \sim 6\times 10^{19} \hbox{ eV}
\left( {B\over 3\times 10^{-6} \hbox{G}} \right)
\left( {D\over 300 \hbox{pc}} \right),\label{Bacc}\end{equation}
with $q_{_M}$ the magnetic charge (inversely proportional to the
electric coupling constant). Using Eq.~(\ref{Bacc}), one gets
table \ref{table} for the maximum energy acquired by a magnetic
monopole (adapted from~\cite{hillas}).

\begin{table}[t]
\begin{center}
\begin{tabular}{*{4}{c}}
\multicolumn{4}{c}{}\\
\multicolumn{4}{c}{\large \bf Magnetic fields, distance and energy}\\ 
\multicolumn{4}{c}{}\\
\hline
object & $B$ (G) & $D$ (pc) & $E_{_M}$ (eV)\\[0.5ex]
\hline
\\[0.5ex]
Neutron Star & $10^{9.5}-10^{13}$ & $10^{-12}$ & $10^{20}-10^{24}$
\\[0.5ex]
White Dwarf & $10^{4}-10^{8}$ & $10^{-9}$ & $10^{18}-10^{22}$
\\[0.5ex]
AGN & $10^{3}-10^{4.5}$ & $10^{-3}$ & $10^{23}-10^{24}$
\\[0.5ex]
SNR & $10^{-5}-10^{-4}$ & $10$ & $10^{19}-10^{20}$
\\[0.5ex]
RG lobes & $10^{-5}-10^{-4}$ & $10^{5}$ & $10^{23}-10^{24}$
\\[0.5ex]
\end{tabular}
\vspace{3mm}
\caption{Characteristic values of magnetic fields $B$ and associated
coherence lengths $D$ for astrophysical objects candidates for
accelerating UHECR. The third column gives the kinetic energy range
that can be obtained for a monopole in such a
field.}
\label{table}
\end{center}
\end{table}

Monopoles are however believed not to be bound to small objects such
as stars (See Ref.~\cite{kp} and references therein). Therefore, among
the various candidates presented on Table~\ref{table}, only Active
Galactic Nuclei (AGN) and Radio Galaxy (RG) lobes have a chance to be
the likely accelerators of monopoles. In both cases, the situation is
such that the regions where the magnetic field is important correspond
also to regions where other particles would be accelerated. These
regions are believed to be filled essentially with electromagnetic
radiation together with very low hadronic density (in the sense that
protons in such a medium would dominantly see the radiation and would
otherwise have a mean free path, with respect to proton-proton
interactions, larger than the acceleration zone). In turn, this
radiation is responsible for degrading the accelerated proton
energies, but is far below the threshold for interacting efficiently
with monopoles.  Therefore, the mean free path of the monopoles is
larger than the actual size of the acceleration region. As a result,
one does not expect any kind of cutoff in the injection spectrum.

Finally, one needs to know the expected number of such events. Let us
assume for the sake of the argument that AGN are the most likely
accelerator candidates. To estimate the flux $\Phi_{_{AGN}}$, we set
$\phi_{_{AGN}}$ the number of monopoles emitted per unit time by a
characteristic AGN, $N_{_{AGN}}$ the number of AGNs up to a distance
which we arbitrarily assume corresponds to a redshift $z=1$ and
$\langle D \rangle$ the mean distance to the AGN under
consideration. One then obtains
\begin{equation}
\Phi_{_{AGN}}=\phi_{_{AGN}}N_{_{AGN}}\langle D \rangle^{-2}.
\end{equation}
The quantity $\phi_{_{AGN}}$ is
\begin{equation}
\phi_{_{AGN}}=\epsilon \, t_0^{-1} N_{_M},
\label{phig}
\end{equation}
with $\epsilon$ the ratio between the power used to accelerate
monopoles and the electromagnetic luminosity, i.e., a measure of the
efficiency of our mechanism, while $t_0$ is the characteristic time
for emitting a monopole. $N_{_M}$ is the number of monopoles present
in the accelerating zone of the AGN. Typically $N_{_M} = \alpha n_{_M}
L_{_{AGN}}^3$, where $\alpha$ is the ratio between the hadronic
($\rho_b$) and monopole ($\rho_{_M}$) densities; $n_{_M}$ is the mean
number density of monopoles in the universe and $L_{_{AGN}}$ the
characteristic size of the acceleration region. Standard monopole
formation mechanisms give~\cite{preskill}
\begin{equation}
n_{_M} \simeq  10^{-19}
\Bigl({m_{_M} \over 10^{11} \hbox{Gev}}\Bigr)^3\quad \hbox{cm}^{-3},
\end{equation}
which implies
\begin{equation}
N_{_M} \simeq \,10^{26}\alpha\Bigl({m_{_M} \over 10^{11}
\hbox{Gev}}\Bigr)^3\,
\Bigl({L_{_{AGN}} \over 10^{-3} \hbox{pc}}\Bigr)^3.
\label{nmono}
\end{equation}
Considering the monopoles to be relativistics so that their velocity
is essentially that of light, the escape time $t_0$ can be assumed to
be the size $L_{_{AGN}}$, so that Eqs. (\ref{nmono}) and (\ref{phig})
combine into
\begin{equation}
\phi_{_{AGN}}\simeq
10^{21}\epsilon\,\alpha\Bigl({m_{_M}
\over 10^{11}\hbox{Gev}}\Bigr)^3
\Bigl({L_{_{AGN}} \over 10^{-3} \hbox{pc}}\Bigr)^2 \quad \hbox{s}^{-1}.
\label{phig2}
\end{equation}

The sphere with $z=1$ contains a baryonic mass $M_{_B} = 5.8\times
10^{54} h^{-1}$g, with $h$ the Hubble constant $H_0$ in units of 75
km$\cdot$s$^{-1}$Mpc$^{-1}$. The total number of galaxies in such a
radius is therefore $N_g = 3\times 10^{10} h^{-1} M_{10}^{-1}$,
$M_{10}$ being the mass of the galaxy in units of $10^{10}$ solar
masses, with $N_{_{AGN}}$ approximately a tenth of this
value~\cite{surveys}. The expected flux $\Phi_{_M}$ is now given by
\begin{equation} \Phi_{_{AGN}} \simeq 3\times 10^{19} \epsilon\,
\alpha_6\, M_{10}\, m_{11}^3\, L_{-3}^2\, \Bigl({\langle D\rangle\over
1\,\hbox{Gpc}}\Bigr)^{-2} \hbox{cm}^{-2}\cdot
\hbox{s}^{-1},\label{flux}\end{equation} with $\alpha_6=\alpha/10^6$,
$m_{11}=m_{_M}/10^{11} {\rm GeV}$ and $L_{-3} = L_{_{AGN}}/10^{-3}
{\rm pc}$. In Eq.~(\ref{flux}), it should be remarked that the
efficiency $\epsilon$ can exceed unity and in fact presumably depends
on the energy scale $\eta$.

\section{Conclusions}

Monopoles represent the most generic prediction of GUT models and as
such are always presented as a possible cosmological nuisance. This is
because they are usually postulated to form at the GUT phase
transition, in which case their remnant density in the universe would
be proportional to the fourth power of this energy
scale~\cite{preskill}, a density considerably larger than the critical
density today. Going back to the original idea of Kephart and
Weiler~\cite{kp}, we consider instead their usefulness in explaining
the mystery of the UHECRs by assuming the simplest of all solution to
the monopole problem, namely that they are produced at a phase
transition taking place at a temperature scale no higher than $10^{9}$
GeV, implying a monopole mass $m_{_M}\alt 10^{11}$ GeV. In this case,
one can safely consider that magnetic monopoles do exist in
reasonnable number in the universe and their study becomes less
academic.

Monopoles as such can hardly interact with air protons to yield air
showers as they must be topologically stable: their expected signal
would be an extremely difficult to observe \v{C}erenkov
shower~\cite{kp}. However, they are obviously easily accelerated to
energies much higher even than the world record observed until now of
$E_{_{W.R.}}\sim 3\times 10^{20}$ eV~\cite{EWR}. One is therefore
tempted to consider them as candidates for explaining those data. The
point we want to make here is that a possibility is left opened when
one considers the nature of monopole.

A monopole is a solitonic configuration of a winding localised Higgs
field. In most reasonnable theories, such a Higgs field not only
permit the symmetry breaking, and thus the appearance of topological
defects, but serves also as a means to provide masses to various
particles to which they couple. This in turn implies that these
particles might get trapped in the monopole core, forming bound states
that might interact with the air protons in such a way as to be
effectively expelled from the monopole. Moreover, they behave as a
neutron in a nucleus: they are stable in the form of bound states, but
unstable otherwise. The observations can then be explained in
supposing that monopoles regularly hit the earth atmosphere with
tremendous energies, releasing only part of it in such ``ionising''
processes. The unstable particles thus obtained are relativistic,
although not highly with a Lorentz fact not exceeding $10^7$ and can
initiate air showers.

We have developped a model in which, to make things simple, the bound
states are formed by means of the condensation of a scalar field. Such
a field can be charged, a point that would modify our analysis by
corrections of order $e^2 \sim 1/137$, but it should be reminded that
the leading effects we studied come from the topological defect itself
and are thus of order $e^{-2}\sim 137$; the correction we neglected is
therefore some four orders of magnitude smaller. Given this
approximation, we have examined in detail the internal structure of
the charged monopole and derived the expected energy levels that might
hopefully be measured in a precise UHECR experiment, as the latter
should, in our model, yield a line spectrum. If such an observation
was done, the formalism we have presented would allow a
straightforward computation of the free parameters and a
reconstruction of the underlying theory. Whether or not the Pierre
Auger Observatory~\cite{PAO} project will fulfill this task is yet an
open question and depends essentially on the spacing of the energy
levels; for some cases, one expects that it will.

Our main result is that such a model satisfies all the present
observational constraints: their expected flux is of the correct order
of magnitude and they interact strongly with the atmosphere. This last
feature comes from the presence of bound states and the Callan-Rubakov
effect~\cite{cr}. A model based on such topological defects would have
the advantages of the non-acceleration scenarios (bottom-up models)
since they would easily propagate, together with the advantages of
acceleration mechanisms, as they need point sources in order to be
observed. Note that this is compatible with the most recent data
(observed doublets and triplets of events within
2.5$^\circ$)~\cite{doublets} implying localized acceleration. These
very data render the previous mechanism for UHECR using
monopoles~\cite{kp} very unprobable.

\section*{Acknowledgments}

We should like to thank S.~Bonazzola, B.~Carter, J.~Dubau, R.~Hakim,
M.~Lemoine, and H.~D.~Sivak for many stimulating and enlightning
discussions.

\end{document}